\documentclass[twocolumn,preprintnumbers,amsmath,amssymb,superscriptaddress]{revtex4}
\usepackage{array}
\usepackage{booktabs}
\usepackage{tabu}
\usepackage{dcolumn}
\usepackage{amsmath}
\usepackage{amsfonts}
\usepackage{amssymb}
\usepackage{graphicx}
\usepackage[colorlinks,linkcolor=red,citecolor=blue]{hyperref}
\usepackage{subfigure}
\usepackage{graphicx}
\usepackage{dcolumn}
\usepackage{bm}
\usepackage{dsfont}

\newcommand{\commentold}[1]{}
\DeclareMathSymbol{:}{\mathpunct}{operators}{"3A}

\def\be{\begin{equation}}
\def\ee{\end{equation}}
\def\bea{\begin{eqnarray}}
\def\eea{\end{eqnarray}}
\def\f{\frac}
\def\n{\nonumber}
\def\l{\label}

\begin{document}
\title{On the contribution of work or heat in exchanged energy via interaction in open bipartite quantum systems}
\author{B. Ahmadi}
\email{b.ahmadi19@gmail.com}
\address{International Centre for Theory of Quantum Technologies, University of Gdansk, Jana Bażyńskiego 1A, 80-309 Gdansk, Poland}
\author{S. Salimi}
\address{Department of Physics, University of Kurdistan, P.O.Box 66177-15175, Sanandaj, Iran}
\author{A. S. Khorashad}
\address{Department of Physics, University of Kurdistan, P.O.Box 66177-15175, Sanandaj, Iran}
\date{\today}

\def\br{\biggr}
\def\bl{\biggl}
\def\Br{\Biggr}
\def\Bl{\Biggl}
\def\be\begin{equation}
\def\ee{\end{equation}}
\def\bea{\begin{eqnarray}}
\def\eea{\end{eqnarray}}
\def\f{\frac}
\def\n{\nonumber}
\def\l{\label}
\begin{abstract}
In this paper, unambiguous redefinitions of heat and work are presented for quantum thermodynamic systems. We will use genuine reasoning based on which Clausius originally defined work and heat in establishing thermodynamics. The change in the energy which is accompanied by a change in the entropy is identified as heat, while any change in the energy which does not lead to a change in the entropy is known as work. It will be seen that quantum coherence does not allow all the energy exchanged between two quantum systems to be only of the heat form. Several examples will also be discussed. Finally, it will be shown that these refined definitions will strongly affect the entropy production of quantum thermodynamic processes giving new insight into the irreversibility of quantum processes.
\end{abstract}

\maketitle
\newpage
\section{Introduction}
\begin{figure}[h]
\centering
\includegraphics[width=5.5cm]{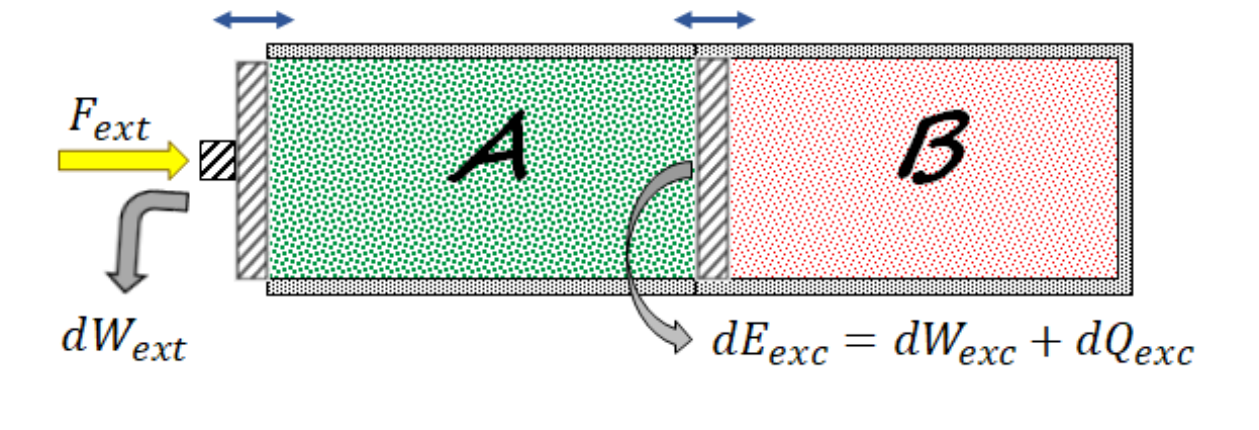}
\caption{(Color online) A classical gas $\mathcal{A}$ (system of interest) is in contact with another gas $\mathcal{B}$ (environment). The total system $\mathcal{AB}$ is insulated against heat from the surroundings. If the pressures of the gases are different from each other and the membrane is movable then work can be done on system $\mathcal{A}$ by system $\mathcal{B}$ through the membrane.}
\label{Fig1}
\end{figure}

In the last few decades we have been witnessing a constantly growing interest in understanding thermodynamic phenomena at the quantum scale \cite{Spohn,Alicki,Gemmer,Bera,Ahmadi,Dolatkhah,Ahmadi1}. Novel fundamental questions arise, such as: how do the laws of thermodynamics emerge in this regime? How can the concepts of heat and work be extended from classical thermodynamics to the quantum realm? How are thermodynamic processes affected by the presence of quantum coherence and entanglement? Extending work and heat from classical thermodynamics to quantum thermodynamics has been one of the major issues in the literature. As is discussed in the following some difficulties appear in identifying work and heat properly that need to be taken care of. In classical thermodynamics a change in the energy of a system is divided into two parts: heat and work \cite{Blundell,Kondepudi,Groot,Yunus},
\begin{equation}\label{4aa}
dE_{\mathcal{A}}=dQ_{\mathcal{A}}+dW_{\mathcal{A}},
\end{equation}
where $dQ_{\mathcal{A}}$ is the heat absorbed by system $\mathcal{A}$ and $dW_{\mathcal{A}}$ the work performed on system $\mathcal{A}$. Eq. (\ref{4aa}) is referred to as the first law of thermodynamics. Heat is defined as the energy in \textit{transit}, between two systems, which is accompanied by a change in the entropy of the system \cite{Blundell,Kondepudi,Groot,Yunus}. And work is defined as the energy in transit which does not lead to any change in the entropy of the system. Heat can only be transferred to the system of interest from another system (environment) through some interaction, while work can be done on the system in two ways: by an external force (field) or by another system via interaction \cite{Blundell,Kondepudi,Groot,Yunus} and since interactions are not generally under the control of the observer therefore some ambiguities may arise in distinguishing work from heat in both classical and quantum setups. For instance, consider a classical gas $\mathcal{A}$ (system of interest) in contact with another classical gas $\mathcal{B}$ with a membrane separating them (see Fig. (\ref{Fig1})). The total system $\mathcal{AB}$ is insulated against heat from the surroundings. Work is done on system $\mathcal{A}$ through the external force, $F_{ext}$, and heat can be transferred to system $\mathcal{A}$ through the membrane. If the membrane is movable, work can also be done on system $\mathcal{A}$ by system $\mathcal{B}$ via the membrane. This means that the exchanged energy between the two systems can be of both heat and work forms, i.e., $dE_{exc}=dW_{exc}+dQ_{exc}$.
\newline
Thus for system $\mathcal{A}$ one has
\begin{equation}\label{4aa1}
dW_{\mathcal{A}}=dW_{ext}+dW_{exc},
\end{equation}
\begin{equation}\label{4aa2}
dQ_{\mathcal{A}}=dQ_{exc}.
\end{equation}
Since the displacement of the membrane is not controlled, $dW_{exc}$ cannot be easily distinguished from $dQ_{exc}$, therefore some ambiguities may arise in identifying work and heat. One may claim that $dW_{exc}$ is not of importance to the observer therefore there is no point in distinguishing $dW_{exc}$ from $dQ_{exc}$. But, as we will see in the following, distinguishing $dW_{exc}$ from $dQ_{exc}$ becomes crucial when investigating the entropy production and the irreversibility of a thermodynamic process \cite{Blundell,Kondepudi,Groot,Yunus}. In classical systems one usually fixes the membrane not to move hence all the exchanged energy is of the heat form, i.e., $dE_{exc}=dQ_{exc}$. However, in the quantum version of the above example usually there is no way to control the interaction between the two quantum systems, hence exchanging some of the energy via the interaction in the form of work is inevitable and it should be carefully taken into account in the definition of work. The aim of this paper is to investigate this issue for quantum thermodynamic systems in exact detail and show how it gives new insight into the irreversibility of a quantum thermodynamic process.
\newline
In 1979, for the first time, R. Alicki defined the heat and work concepts for quantum thermodynamic systems in the weak coupling limit \cite{Alicki}. It was assumed that a change in the local Hamiltonian of a system is necessarily associated with work and any change in the state of a system is necessarily associated with heat. Due to this association between the change in the local Hamiltonian of the system and work only the contribution of the external field is considered in the definition of work. In fact, the work done by the environment on the system, through interaction, is taken into account as heat which is not reasonable. In view of this approach, for a system $\mathcal{A}$ work and heat are defined, respectively, as \cite{Alicki}
\begin{equation}\label{4a}
d\langle W_{\mathcal{A}}(t)\rangle\equiv tr\{\rho_{\mathcal{A}}(t)dH_{\mathcal{A}}(t)\},
\end{equation}
\begin{equation}\label{4}
d\langle Q_{\mathcal{A}}(t)\rangle\equiv tr\{d\rho_{\mathcal{A}}(t)H_{\mathcal{A}}(t)\}.
\end{equation}
As can be seen from Eq. (\ref{4a}) only the change in the Hamiltonian of system $\mathcal{A}$ contributes to the work done on the system and the work done by the environment, through the interaction, has not been considered in the definition of work. Although such definitions have been widely accepted within the context of quantum thermodynamics and seem to directly satisfy the first law of thermodynamics we will show that there exists some ambiguities in these definitions. We will show that part of the heat, defined in Eq. (\ref{4}), is in fact the work done on the system by the environment through the interaction. Therefore unambiguous redefinitions of work and heat are needed for quantum thermodynamic systems. In order to do this we will use the genuine reasoning based on which Clausius defined work and heat, in the first place, to establish thermodynamics. With our new and novel definitions of work and heat all the problems mentioned above are resolved. It will also be seen that quantum coherence appears to be a resource for doing work by the system.
\section{Work and heat in quantum thermodynamics}
\begin{figure}[h]
\centering
\includegraphics[width=4cm]{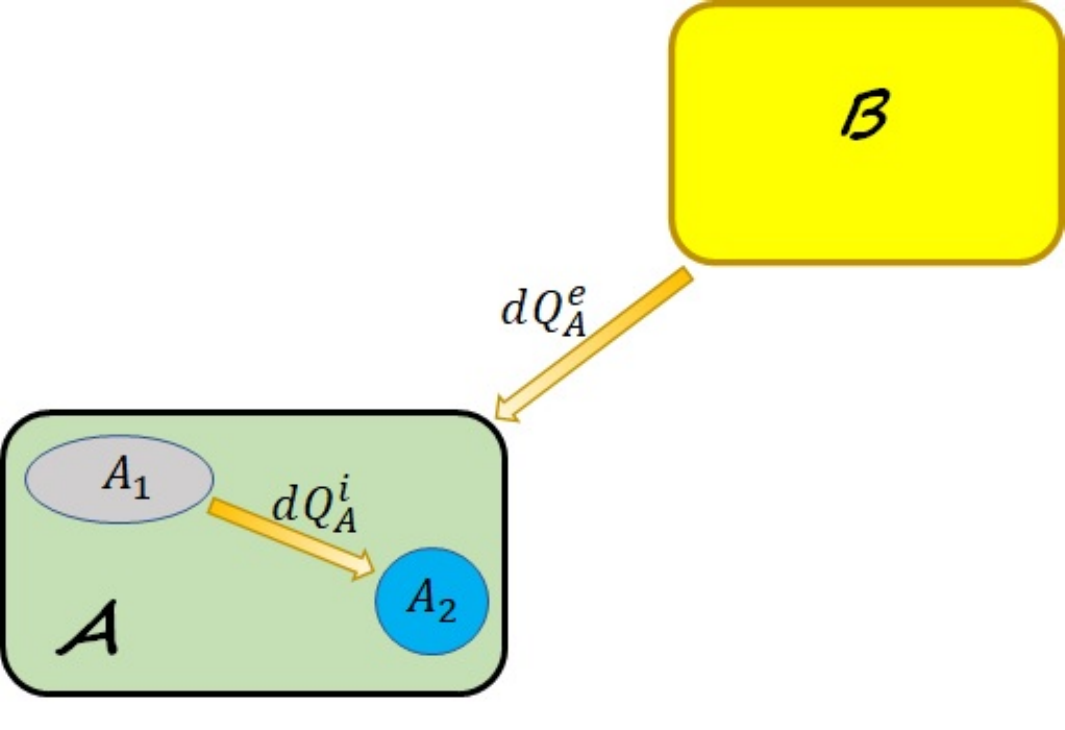}
\caption{(Color online) System $\mathcal{A}$ interacting with system $\mathcal{B}$. $dQ^e_{\mathcal{A}}$ is the heat flow, in the exterior of system $\mathcal{A}$ due to the interaction with system $\mathcal{B}$, which contributes to $d_eS_{\mathcal{A}}$ and $dQ^i_{\mathcal{A}}$ is the heat flow, in the interior of system $\mathcal{A}$, which contributes to $d_iS_{\mathcal{A}}$.}
\label{Fig2}
\end{figure}

\begin{figure}[h]
\centering
\includegraphics[width=5.5cm]{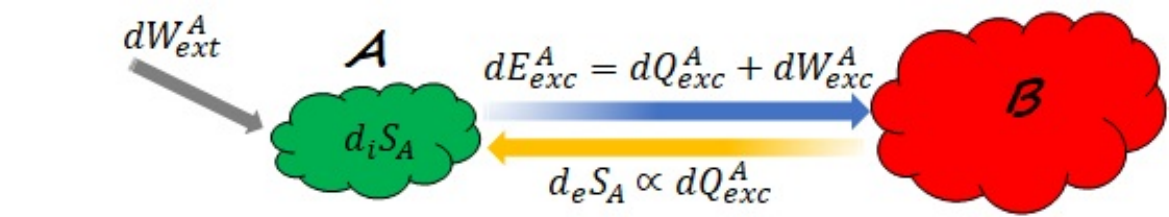}
\caption{(Color online) A quantum system $\mathcal{A}$ (system of interest) is in contact with another quantum system $\mathcal{B}$ (environment). The total system $\mathcal{AB}$ is insulated against heat from the surroundings. There is flow of energy between the two systems due to the interaction. External work is also done on system $\mathcal{A}$ through an external field.}
\label{Fig3}
\end{figure}

In classical thermodynamics using the definition of heat and work the entropy of the system is defined \cite{Blundell,Kondepudi,Groot,Yunus} but in quantum thermodynamics the scenario is the converse, i.e., using the definition of the entropy we define heat and work. Assume that the state of a quantum system $\mathcal{A}$ with Hamiltonian $H_{\mathcal{A}}(t)$ at time $t$ is $\rho_{\mathcal{A}}(t)$ which can always be uniquely decomposed into its instantaneous eigenvectors as \cite{Audretsch}
\begin{equation}\label{1}
\rho_{\mathcal{A}}(t)=\sum_{i=1}^{d}p_i(t)|\psi_i(t)\rangle\langle\psi_i(t)|,
\end{equation}
where $|\psi_i(t)\rangle$ are the eigenvectors of $\rho_{\mathcal{A}}(t)$ and $p_i(t)$ the corresponding occupation of probabilities at time $t$. The Von Neumann entropy of a quantum state $\rho_{\mathcal{A}}(t)$ is defined as \cite{Nielsen,Breuer}
\begin{equation}\label{2}
S(\rho_{\mathcal{A}}(t))=-tr\{\rho_{\mathcal{A}}(t)\ln\rho_{\mathcal{A}}(t)\}=-\sum_{i=1}^{d}p_i(t)\ln p_i(t).
\end{equation}
Thus the infinitesimal change in the entropy reads
\begin{equation}\label{2a}
dS(\rho_{\mathcal{A}}(t))=-\sum_{i=1}^{d}dp_i(t)\ln p_i(t).
\end{equation}
Clausius showed that the total change in the entropy of any thermodynamic system may be divided into two completely different parts as \cite{Blundell,Kondepudi,Groot,Yunus}
\begin{equation}\label{2aa}
dS_{\mathcal{A}}=d_eS_{\mathcal{A}}+d_iS_{\mathcal{A}},
\end{equation}
where $d_eS_{\mathcal{A}}$ is the flow of information caused by the flow of heat in the "exterior" of the system, i.e., $d_eS\propto dQ^e_{\mathcal{A}}$ and $d_iS_{\mathcal{A}}$ is the entropy produced due to the irreversible flow of heat in the "interior" of the system, i.e., $d_iS_{\mathcal{A}}\propto dQ^i_{\mathcal{A}}$ (see Fig. (\ref{Fig2})). This division is very crucial and it must be carefully taken into account in defining work and heat, therefore we will refer to it in defining work and heat for quantum thermodynamic processes. It will be seen that this important point has been ignored for the definitions of work and heat introduced in Eqs. (\ref{4a}) and (\ref{4}).
\newline
Now using Eq. (\ref{1}) the infinitesimal change in $\rho_{\mathcal{A}}(t)$ can be expressed as
\begin{eqnarray}\label{3}\nonumber
d\rho_{\mathcal{A}}(t)&=&\sum_{i=1}^{d}dp_i(t)|\psi_i(t)\rangle\langle\psi_i(t)|\\
&+&\sum_{i=1}^{d}p_i(t)d(|\psi_i(t)\rangle\langle\psi_i(t)|).
\end{eqnarray}
Thus as can be seen from Eq. (\ref{3}) the change in the state of the system is divided into two parts: the first part is due to the change in $p_i(t)$, which is caused by the non-unitary part of the dynamics, and the second part due to the change in the eigenvectors of the state, which is caused by the unitary part of the dynamics. The former will lead to a change in the entropy of the state but the latter will not. This can clearly be seen by comparing Eq. (\ref{2a}) with Eq. (\ref{3}). The average internal energy of a quantum system at time $t$ is defined as \cite{Gemmer}
\begin{equation}\label{3a}
\langle E_{\mathcal{A}}(t)\rangle=tr\{\rho_{\mathcal{A}}(t)H_{\mathcal{A}}(t)\}.
\end{equation}
Therefore the change in the internal energy reads
\begin{equation}\label{3aa}
d\langle E_{\mathcal{A}}(t)\rangle=tr\{d\rho_{\mathcal{A}}(t)H_{\mathcal{A}}(t)\}+tr\{\rho_{\mathcal{A}}(t)dH_{\mathcal{A}}(t)\}.
\end{equation}
Using Eq. (\ref{3}) the first term on the RHS of Eq. (\ref{3aa}) becomes
\begin{eqnarray}\label{5}\nonumber
tr\{d\rho_{\mathcal{A}}(t)H_{\mathcal{A}}(t)\}&=&tr\{\sum_{i=1}^{d}dp_i(t)|\psi_i(t)\rangle\langle\psi_i(t)|H_{\mathcal{A}}(t)\}\\ \nonumber
&+&tr\{\sum_{i=1}^{d}p_i(t)d(|\psi_i(t)\rangle\langle\psi_i(t)|)H_{\mathcal{A}}(t)\},\\
\end{eqnarray}
which is the energy exchanged between the system and the environment through the interaction (see Fig. (\ref{Fig3})). As was mentioned above heat is part of the exchanged energy (between two systems) which leads to a change in the entropy of the system and that part of the exchanged energy which does not lead to any change in the entropy of the system is considered to be the work done on the system. Therefore the second term on the RHS of Eq. (\ref{5}) should be considered as work, $dW^\mathcal{A}_{exc}$, rather than heat because it comes from the unitary part of the evolution (see Appendix A) and consequently it does not lead to any change in the entropy of system $\mathcal{A}$. The microscopic decomposition of the exchanged energy, in Eq. (\ref{5}), into two parts is a new unraveling of the first law of thermodynamics for quantum systems that constitutes one of our main results (see Fig. (\ref{Fig3})). For the second term on the RHS of Eq. (\ref{3aa}) we have
\begin{equation}\label{6}
tr\{\rho_A(t)dH_A(t)\}=tr\{\sum_{i=1}^{d}p_i(t)|\psi_i(t)\rangle\langle\psi_i(t)|dH_A(t)\}.
\end{equation}
Now using Eqs. (\ref{5}) and (\ref{6}) the total change in the internal energy of the system becomes
\begin{eqnarray}\label{7}\nonumber
d\langle E_{\mathcal{A}}(t)\rangle&=&tr\{\sum_{i=1}^{d}dp_i(t)|\psi_i(t)\rangle\langle\psi_i(t)|H_{\mathcal{A}}(t)\}\\ \nonumber
&+&tr\{\sum_{i=1}^{d}p_i(t)d(|\psi_i(t)\rangle\langle\psi_i(t)|)H_{\mathcal{A}}(t)\}\\
&+&tr\{\sum_{i=1}^{d}p_i(t)|\psi_i(t)\rangle\langle\psi_i(t)|dH_{\mathcal{A}}(t)\}.
\end{eqnarray}
As was discussed above, only the first term on the RHS of Eq. (\ref{7}) leads to a change in the entropy of the system hence this term is to be considered as heat. The second and third terms are the work done on the system. In other words, only the first part of the energy change takes away (or brings) information from (into) the system. Accordingly the appropriate definitions of heat and work for a quantum thermodynamic system $\mathcal{A}$, respectively, are
\begin{equation}\label{8}
d\langle Q_{\mathcal{A}}(t)\rangle\equiv tr\{\sum_{i=1}^{d}dp_i(t)|\psi_i(t)\rangle\langle\psi_i(t)|H_{\mathcal{A}}(t)\},
\end{equation}
\begin{equation}\label{9}
d\langle W_{\mathcal{A}}(t)\rangle\equiv tr\{\sum_{i=1}^{d}p_i(t)d(|\psi_i(t)\rangle\langle\psi_i(t)|H_{\mathcal{A}}(t))\}.
\end{equation}
\newline
It must be mentioned that one may object that the second term on the RHS of Eq. (\ref{7}) can still be considered as heat even if it does not lead to any change in the entropy of the system because there might be the same term in $\Delta_iS_{\mathcal{A}}$ with the opposite sign such that they cancel out each other consequently the second term on the RHS of Eq. (\ref{7}) is not accompanied with any change in the entropy of the system. The problem with this reasoning is that the second term on the RHS of Eq. (\ref{7}) is part of the energy flow between the system and the environment, i.e., the energy flow in the exterior of the system hence, as we mentioned earlier, it cannot contribute anything to the entropy produced in the interior of the system. What is of particular interest about these new definitions in Eqs. (\ref{8}) and (\ref{9}) is that even if the Hamiltonian $H_{\mathcal{A}}(t)$ remains unchanged work can still be done on the system through the interaction while according to Alicki's definitions it is zero in this case. This is plausible because when two systems with constant Hamiltonian $H=H_{\mathcal{A}}+H_{\mathcal{B}}+H_{\mathcal{AB}}$ are interacting work is expected to be done on both systems through the interaction Hamiltonian $H_{\mathcal{AB}}$. This is especially important whenever an external field is replaced by a quantized degree of freedom for which the time dependence of the Hamiltonian is removed at the expense of increasing the dimensionality of the Hamiltonian. Let us now examine how coherence of a state is related to doing work on the system. Eq. (\ref{3aa}) can also be written in the form \cite{Kieu,Quan}
\begin{equation}\label{10}
d\langle E_{\mathcal{A}}(t)\rangle=\sum_{i=1}(E_i(t)dq_i(t)+q_i(t)dE_i(t)),
\end{equation}
in which $E_i(t)$ is the ith eigenenergy of the quantum system $\mathcal{A}$ at time $t$ with the time-dependent Hamiltonian $H_{\mathcal{A}}(t)=\sum_{i=1}E_i(t)|E_i(t)\rangle\langle E_i(t)|$ and $q_i(t)$ the probability of the system to be in the eigenstate $|E_i(t)\rangle$ at time $t$. Analogous to Alicki's definition the following identification was made to define heat and work \cite{Kieu,Quan},
\begin{equation}\label{11}
d\langle Q_{\mathcal{A}}(t)\rangle\equiv \sum_{i=1}E_i(t)dq_i(t),
\end{equation}
\begin{equation}\label{12}
d\langle W_{\mathcal{A}}(t)\rangle\equiv \sum_{i=1}q_i(t)dE_i(t).
\end{equation}
Since $q_i(t)$ in Eq. (\ref{10}) are different from $p_i(t)$ in Eqs. (\ref{1})-(\ref{3}) then the change in $q_i(t)$ does not necessarily lead to a change in the entropy of the system (see Appendix B for more details). Therefore defining heat as in Eq. (\ref{11}) is not suitable. The change in $q_i(t)$ will necessarily lead to a change in the entropy of the system only when the state of the system is diagonal in the energy eigenbasis at any time $t$. This means that coherence, with respect to the energy eigenbasis, plays an important role in proper definitions of heat and work in quantum thermodynamics. When coherence of a state changes the eigenvectors of the state also change then according to Eq. (\ref{9}) this, in turn, will lead to doing work.
\section{Examples}
\begin{figure}[h]
\centering
\includegraphics[width=5.5cm]{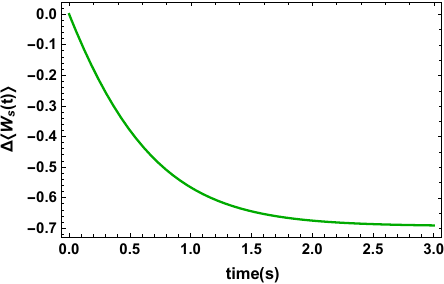}
\caption{(Color online) The work $\Delta\langle W_s(t)\rangle$ vs time $t$ for a two-level system in contact with a thermal bath with decay rate $\gamma(t)=0.2$ and $\omega_0=2$. As can be seen work is done on the system by the bath through the interaction and as the system approaches equilibrium the bath stops doing work on the system.}
\label{Fig4}
\end{figure}

In order to illustrate the difference between our definitions of heat and work and the definitions introduced in Eqs. (\ref{4a}) and (\ref{4}) the following examples are discussed. Let us first examine the case of the interaction of an atom with a field \cite{Gerry}. The field could be considered to be classical or fully quantized. We first turn to the case where an atom is driven by a classical sinusoidal electric field. We assume that the field has the form $\textbf{E}(t)=\textbf{E}_0\cos(\omega t)$, $\omega$ being the frequency of the radiation. Thus the Hamiltonian becomes \cite{Gerry}
\begin{equation}\label{14}
H(t)=H_{atom}-\textbf{d}.\textbf{E}(t),
\end{equation}
where $\textbf{d}$ is the dipole moment operator of the atom. Since the Hamiltonian is time-dependent Eqs. (\ref{4a}) and (\ref{9}) are both nonzero. But if the field is treated fully quantized the total Hamiltonian reads \cite{Gerry}
\begin{equation}\label{15}
H=H_{atom}+H_{field}+H_I,
\end{equation}
where $H_{field}=\hbar\omega a^\dagger a$ and $H_I=-\textbf{d}.\mathcal{E}_0(a-a^{\dagger})$ and $\mathcal{E}_0$ is a constant vector. In this case Eq. (\ref{4a}) equals zero, i.e., no work is extracted by the field but using Eq. (\ref{9}) work is clearly extracted from the atom which is expected. For the last example consider a two-level (spin-1/2) system $S$ interacting with a thermal bath of harmonic oscillators at temperature $T$ \cite{Breuer}. The total Hamiltonian of the system and the bath reads
\begin{equation}\label{13}
H=H_S+H_B+H_{SB},
\end{equation}
in which $H_S=(\omega_0/2)\sigma_z$ is the free Hamiltonian of the system with $\omega_0>0$ the transition frequency and $\sigma_z$ the Pauli matrix, $H_B=\sum_{i}\omega_ia^\dag(\omega_i) a(\omega_i)$ the Hamiltonian of the bath and $H_{SB}=\sum_{i}g(\omega_i)(\sigma_-a^\dag(\omega_i)+\sigma_+a(\omega_i))$ the interaction Hamiltonian with $g(\omega_i)$ the coupling strength and $\sigma_{\pm}=(\sigma_x\pm i\sigma_y)/2$. We consider the dynamics to be Markovian therefore the coupling is weak and the stationary solution of the master equation is equal to the thermal equilibrium state $\rho_s^{th}=\exp(-\beta H_s)/Z_s$ where $\beta=1/T$. If we choose, for example, the system to be initially in the ground state, i.e., $\rho_s(0)=\begin{pmatrix}
  0 & 0 \\  0 & 1
\end{pmatrix}$ then the eigenvectors of the state remains unchanged throughout the whole evolution and since the free Hamiltonian of the system $H_S$ is constant thus using Eq. (\ref{9}) no work is done on the system, i.e., all the energy exchanged between the system and the thermal bath occurs in the form of heat. Therefore, in this specific case the results are the same for both Alicki's and our frameworks. But if the initial state of the system is $\rho_s(0)=\dfrac{1}{2}\begin{pmatrix}
  1 & 1 \\  1 & 1
\end{pmatrix}$, i.e., the initial state contains coherence with respect to the energy eigenbasis, then the eigenvectors of the state of the system keeps varying until the state reaches equilibrium which has no coherence. Hence based on Eq. (\ref{9}), due to varying eigenvectors of the state, work done on the system is not zero in this case (see illustration in Fig. (\ref{Fig4})). As depicted in Fig. (\ref{Fig4}), for the initial state $\rho_s(0)=\dfrac{1}{2}\begin{pmatrix}
  1 & 1 \\  1 & 1
\end{pmatrix}$, the environment does work on the system, through the interaction, from the beginning until the system reaches thermal equilibrium with the bath.

\begin{figure}[h]
\centering
\includegraphics[width=5.5cm]{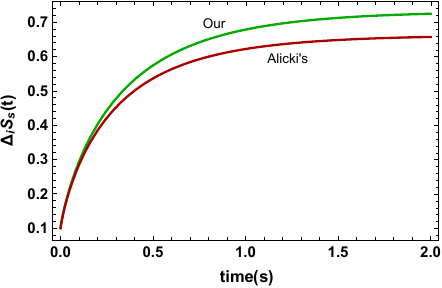}
\caption{(Color online) The entropy production of the system $\Delta_iS_s(t)$ vs time $t$ for a two-level system with $\omega_0=2$ in contact with a thermal bath with temperature $T=10K$ and decay rate $\gamma(t)=0.2$. It is seen that removing that part of the energy, which does not play any role in the change of the entropy of the system, from the definition of the heat reveals that more irreversibility is, in fact, occurring inside the system.}
\label{Fig6}
\end{figure}
As was mentioned before, this example implies that the initial coherence of the state of the system contributes to doing work. In other words, "coherence contained in the state of the system does not allow the energy exchanged between the system and the environment to be only of the heat form". In further publications we will investigate the connection between the initial coherence and the exchanged work in more detail. Initial coherence is of great significance especially in charging a quantum battery by a quantum charger because we need to extract all the energy transferred to the battery in the form of work \cite{Binder,Campaioli,Ferraro,Barra,Andolina}. In Ref. \cite{Andolina} it was shown that if the initial state of the charger is a coherent state the extractable work from the battery using a cyclic unitary transformation is optimal. It has also been shown that the extractable work in fully quantized setups obtainable from non-passivity strongly depends on the initial state of the system, particularly on its coherence \cite{Klimovsky}. Finally, we will show that these refined definitions of work and heat will give new insight into the irreversibility and the entropy production of a quantum thermodynamic system. The entropy production of a thermodynamic system $\mathcal{A}$ is defined as \cite{Blundell,Kondepudi,Groot,Yunus}
\begin{equation}\label{14}
\Delta_iS_\mathcal{A}=:\Delta S_\mathcal{A}-\dfrac{\Delta Q_\mathcal{A}}{T},
\end{equation}
where $T$ is the temperature of the bath with which the system interacts. As can be seen from Eq. (\ref{14}) the entropy production of a system depends on the definition of heat, therefore since the entropy production is a measure of irreversibility of a thermodynamic process these new refined definitions of heat and work give new insight into irreversibility of quantum thermodynamic processes. In Fig. (\ref{Fig6}) the entropy production of the system, investigated in the last example for the initial state $\rho_s(0)=\dfrac{1}{2}\begin{pmatrix}
  1 & 1 \\  1 & 1
\end{pmatrix}$, is illustrated. As is shown the entropy production in Alicki's framework is less than that in our framework. This means that removing that part of the energy, which does not play any role in the change of the entropy of the system, from the definition of heat reveals that more irreversibility is, in fact, occurring in the interior of the system. Therefore our novel refined definitions of heat and work give new insight into the irreversibility of quantum thermodynamic processes. It should be mentioned that in Refs. \cite{Boukobza,Weimer} different frameworks form Alicki's were proposed. But in both frameworks the change in the eigenvectors of the state was never taken into account.
\newline
Here it must be pointed out that R. Alicki's definitions of work and heat, and therefore our refined definitions, are only defined for the systems interacting in the weak coupling limit where there exist no correlations between the system of interest with the environment \cite{Alicki}. As an example in which these definitions fail consider the evolution of a composite system $\mathcal{AB}$. The interaction is such that $[H_\mathcal{A}\otimes\mathbb{I}_\mathcal{B}, H_{int}]=0$ where $H_\mathcal{A}=(\omega_0/2)\sigma_z$ and $H_{int}=\lambda\sigma_z\otimes\sigma_z$ (dephasing). $\lambda$ is considered to be small enough such that the interaction energy is negligible compared to internal energy of subsystems. With the initial states $\rho_\mathcal{A}(0)=\begin{pmatrix}
  p & c \\  c^* & 1-p
\end{pmatrix}$ and $\rho_{\mathcal{B}}(0)=\begin{pmatrix}
  1/2 & 0 \\  0 & 1/2
\end{pmatrix}$ the dynamics of subsystem $\mathcal{A}$ is obtained as $\rho_\mathcal{A}(t)=\begin{pmatrix}
  p & c\cos2\lambda t \\  c^*\cos2\lambda t & 1-p
\end{pmatrix}$ and the state of subsystem $\mathcal{B}$ remains unchanged. After some straightforward calculations it is seen that the internal energy of subsystem $\mathcal{A}$ remains unchanged during the dynamics and since $H_\mathcal{A}$ is constant based on Alicki's framework no work is done and consequently no heat is transferred. Interestingly, the same scenario occurs according to our framework, i.e., no work is done and no heat is transferred. This seems not to satisfy the first law of thermodynamics because the two subsystems $\mathcal{A}$ and $\mathcal{B}$ are interacting and the change in the entropy of subsystem $\mathcal{A}$ is nonzero
\begin{equation}\label{15A}
dS_{\mathcal{A}}=-\lambda\sin(2\lambda t)\ln\tan^2(\lambda t)\neq0,
\end{equation}
while no work and heat are exchanged. This nonzero $dS_{\mathcal{A}}$ comes from the \textit{correlations} established between the two subsystems. In other words, system $\mathcal{A}$ here only exchanges information through (with!) correlations. Let us clarify this issue further. In this example the state of subsystem $\mathcal{B}$ stays unchanged thus $dS_{\mathcal{B}}=0$. We know that \cite{Nielsen}
\begin{equation}\label{SAB}
dS_{\mathcal{AB}}=dS_{\mathcal{A}}+dS_{\mathcal{B}}+dS_{\mathcal{C}},
\end{equation}
where $dS_{\mathcal{C}}$ is the entropy in the correlations. Since the total system is closed $dS_{\mathcal{AB}}=0$ therefore
\begin{equation}\label{SAC}
dS_{\mathcal{A}}=-dS_{\mathcal{C}}.
\end{equation}
This means that information is leaving (entering) subsystem ${\mathcal{A}}$ to (from) correlations. Generally this is true for interaction Hamiltonians, time-independent or not, which commute with the system Hamiltonian $[H_\mathcal{A}\otimes\mathbb{I}_\mathcal{B}, H_{int}]=0$.
\section{Summary}We derived the proper definitions of work and heat for quantum thermodynamic systems. We then observed that part of the energy exchanged between two quantum systems, due to the interaction, can be in the form of work. This microscopic decomposition of the exchanged energy into heat and work is a new unraveling of the first law of thermodynamics for quantum systems. It was also shown that quantum coherence plays a major role in doing work such that quantum coherence does not allow the exchanged energy between two quantum systems to be only of the heat form. As is expected theses refined definitions of heat and work will strongly affect the entropy production of quantum processes giving new insight into irreversibility of quantum thermodynamic processes.
\section*{Acknowledgment} We acknowledge support from the Foundation for Polish Science through IRAP project co-financed by EU within the Smart Growth Operational Program (contract no.2018/MAB/5). The authors would like to thank F. Kheirandish and G. De Chiara for very useful comments and advises.
\newline
\textit{Note added}.—After completion of this work, we became aware of another independent work with entropy-based separation of energy \cite{Alipour}.

\section*{Appendix A}\label{AppendixA}
The infinitesimal change in the eigenvectors of the state of the system $\rho_\mathcal{A}(t)$ in Eq. (\ref{5}) is achieved as
\begin{eqnarray}\label{A}\nonumber
d(|\psi_i(t)\rangle\langle\psi_i(t)|)&=&|\psi_i(t+dt)\rangle\langle\psi_i(t+dt)|-|\psi_i(t)\rangle\langle\psi_i(t)|\\ \nonumber
&=&U(dt)|\psi_i(t)\rangle\langle\psi_i(t)|U^\dagger(dt)\\
&-&|\psi_i(t)\rangle\langle\psi_i(t)|,
\end{eqnarray}
where $U(dt)$ is the unitary operator transforming the orthogonal basis $\{|\psi_i(t)\rangle\}_{i=1}^d$ to another orthogonal basis $\{|\psi_i(t+dt)\rangle\}_{i=1}^d$ \cite{Heinosaari}. Using Eqs. (\ref{1}) and (\ref{A}) we have
\begin{eqnarray}\label{A1}\nonumber
\sum_{i=1}^{d}p_i(t)d(|\psi_i(t)\rangle\langle\psi_i(t)|)&=&U(dt)\rho_\mathcal{A}(t)U^\dagger(dt)\\
&-&\rho_\mathcal{A}(t).
\end{eqnarray}
Eq. (\ref{A1}) is the unitary part of the total change in the state of the system $\rho_\mathcal{A}(t)$. Now substituting Eq. (\ref{A1}) into Eq. (\ref{5}) the second term on the right hand side of Eq. (\ref{5}) reads
\begin{equation}\label{A2}
tr\{\sum_{i=1}^{d}p_i(t)d(|\psi_i(t)\rangle\langle\psi_i(t)|)H_\mathcal{A}(t)\}=tr\{d\rho_s^U(t)H_\mathcal{A}(t)\},
\end{equation}
in which $d\rho_\mathcal{A}^U(t)\equiv U(dt)\rho_\mathcal{A}(t)U^\dagger(dt)-\rho_\mathcal{A}(t)$. Thus Eq. (\ref{A2}) is in fact the energy which is unitarily exchanged between the two quantum systems through the interaction.
\section*{Appendix B}\label{AppendixB}
The average of the internal energy of system $\mathcal{A}$ at time $t$ is defined as \cite{Gemmer}
\begin{eqnarray}\label{I}\nonumber
\langle E_\mathcal{A}(t)\rangle&=& tr\{\rho_\mathcal{A}(t)\,H_\mathcal{A}(t)\}\\ \nonumber
&=&tr\{\sum_{i=1}^d p_i(t)|\psi_i(t)\rangle \langle\psi_i(t)|H_\mathcal{A}(t)\}\\
&=&\sum_{i=1}^d p_i(t)\langle\psi_i(t)|H_\mathcal{A}(t)|\psi_i(t)\rangle.
\end{eqnarray}
On the other hand
\begin{equation}\label{II}
H_\mathcal{A}(t)=\sum_{j}E_j(t)|E_j(t)\rangle\langle E_j(t)|,
\end{equation}
therefore
\begin{eqnarray}\label{E}\nonumber
\langle E_\mathcal{A}(t)\rangle&=&\sum_{i=1}^d \sum_{j}p_i(t)\,E_j(t)|\langle\psi_i(t)|E_j(t)\rangle|^2\\
&=&\sum_jq_j(t)\,E_j(t),
\end{eqnarray}
where
\begin{eqnarray}\label{pq}\nonumber
q_j (t)&=&\sum_{i=1}^d p_i (t)\,|\langle\psi_i (t)|E_j (t)\rangle|^2,\\
&=&\sum_{i=1}^d p_i (t)\,R_{i\to j} (t),
\end{eqnarray}
with $R_{i\to j}(t)$ the transition probability from the eigenbasis $|\psi_i (t)\rangle$ to the eigenbasis $|E_j (t)\rangle$. The connection between $p_i (t)$ and $q_i (t)$ is given through Eq. (\ref{pq}).
Now from Eqs. (\ref{E}) and (\ref{pq}) we have
\begin{equation}
\langle E_\mathcal{A}(t)\rangle=\sum_{i=1}^d\sum_j p_i (t)\, R_{i\to j} (t)\,E_j (t).
\end{equation}
Then
\begin{eqnarray}\nonumber
d\langle E_\mathcal{A}(t)\rangle&=&\underbrace{\sum_{i=1}^d\sum_jdp_i(t)\,R_{i\to j}(t)\,E_j(t)}_{d\langle Q_\mathcal{A}(t)\rangle}\\
&+&\underbrace{\sum_{i=1}^d\sum_j p_i(t)\,d(R_{i\to j}(t)\,E_j(t))}_{d\langle R(t)\rangle}.
\end{eqnarray}
Hence we have
\begin{equation}
d\langle Q_\mathcal{A}(t)\rangle=\sum_{i=1}^d\sum_jdp_i(t)\,R_{i\to j}(t)\,E_j(t),
\end{equation}
\begin{eqnarray}\nonumber
d\langle W_\mathcal{A}(t)\rangle&=&\sum_{i=1}^d\sum_j p_i(t)\,\textcolor[rgb]{1.00,0.00,0.00}{dR_{i\to j}(t)}\,E_j(t)\\
&+&\sum_{i=1}^d\sum_jp_i(t)\, R_{i\to j}(t)\,\textcolor[rgb]{1.00,0.00,0.00}{dE_j(t)},
\end{eqnarray}
which means that two terms contribute to $dW(t)$, one is originating from the variation of transition probabilities (the first term) and the other originates from the variations of the energy levels (the second term).

\begin{references}
\bibitem{Spohn} H. Spohn, \textcolor{blue}{Journal of Mathematical Physics \textbf{19}, 1227 (1978)}.
\bibitem{Alicki} R. Alicki, \textcolor{blue}{J. Phys. A \textbf{12}, L103 (1979)}.
\bibitem{Gemmer} J. Gemmer, M. Michel, and G. Mahler, Quantum Thermodynamics: Emergence of Thermodynamic Behavior Within Composite Quantum Systems, Lect. Notes Phys. 784 (Springer, Berlin Heidelberg 2009).
\bibitem{Bera} M. N. Bera, A. Riera, M. Lewenstein, and A. Winter, \textcolor{blue}{Nat. Commun. \textbf{8}, 2180 (2017)}.
\bibitem{Ahmadi} B. Ahmadi, S. Salimi, A. S. Khorashad and F. Kheirandish, \textcolor{blue}{Sci. Rep. \textbf{9}, 8746 (2019)}.
\bibitem{Dolatkhah} H. Dolatkhah, S. Salimi, A. S. Khorashad, and  S. Haseli, \textcolor{blue}{Sci. Rep. \textbf{10}, 9757 (2020)}.
\bibitem{Ahmadi1} B. Ahmadi, S. Salimi, and A. S. Khorashad, \textcolor{blue}{\href{https://arxiv.org/abs/2002.10747} {arXiv:1912.01983 (2019)}}.
\bibitem{Blundell} S. J. Blundell and K. M. Blundell, Concepts in Thermal Physics (Oxford University Press 2009).
\bibitem{Kondepudi} D. Kondepudi and I. Prigogine, Modern Thermodynamics (New York: Wiley 1998).
\bibitem{Groot} S. R. De Groot  and P. Mazur, Non-Equilibrium Thermodynamics (New York: Dover, 1984).
\bibitem{Yunus} Yunus A. \c{C}engel and Michael A. Boles,  Thermodynamics: An Engineering Approach, 8th edn (New York, McGraw-Hill 2015)
\bibitem{Audretsch} J. Audretsch, Entangled Systems: New Directions in Quantum Physics (New York: Wiley 2007).
\bibitem{Nielsen}  M. A. Nielsen and I. L. Chuang, Quantum Computation and Quantum Information (Cambridge University Press, 2000).
\bibitem{Breuer} H. P. Breuer and F. Petruccione, The theory of open quantum systems (Oxford University Press, Oxford, 2002).
\bibitem{Kieu} T. D. Kieu \textcolor{blue}{Phys. Rev. Lett. \textbf{93}, 140403 (2004)}.
\bibitem{Quan} H. T. Quan, P. Zhang, and C. P. Sun, \textcolor{blue}{Phys. Rev. E \textbf{72}, 056110 (2005)}.
\bibitem{Gerry} C. Gerry and P. Knight, Introductory Quantum Optics (Cambridge University Press, 2005).
\bibitem{Binder} F. C. Binder, S. Vinjanampathy, K. Modi and J. Goold, \textcolor{blue}{New J. Phys. \textbf{17} 075015 (2015)}.
\bibitem{Campaioli} F. Campaioli, F. A. Pollock, F. C. Binder, L. Céleri, J. Goold, S. Vinjanampathy, and K. Modi, \textcolor{blue}{Phys. Rev. Lett. \textbf{118}, 150601 (2017)}.
\bibitem{Ferraro} D. Ferraro, M. Campisi, G. M. Andolina, V. Pellegrini, and M. Polini, \textcolor{blue}{Phys. Rev. Lett. \textbf{120}, 117702 (2018)}.
\bibitem{Barra} Felipe Barra, \textcolor{blue}{Phys. Rev. Lett. \textbf{122}, 210601 (2019)}.
\bibitem{Andolina} G. M. Andolina, M. Keck, A. Mari, M. Campisi, V. Giovannetti, and M. Polini, \textcolor{blue}{Phys. Rev. Lett. \textbf{122}, 047702 (2019)}.
\bibitem{Klimovsky} D. Gelbwaser-Klimovsky, R. Alicki, and G. Kurizki, \textcolor{blue}{EPL, \textbf{103} 60005 (2013)}.
\bibitem{Boukobza} E. Boukobza and D. J. Tannor, \textcolor{blue}{Phys. Rev. A \textbf{74}, 063823 (2006)}.
\bibitem{Weimer} H. Weimer, M. J. Henrich, F. Rempp, H. Schr\"{o}der and G. Mahler, \textcolor{blue}{EPL, \textbf{83} 30008 (2008)}.
\bibitem{Alipour} S. Alipour,  A. T. Rezakhani, A. Chenu, A. del Campo and T. Ala-Nissila ,  \textcolor{blue}{\href{https://journals.aps.org/pra/abstract/10.1103/PhysRevA.105.L040201} {Phys. Rev. A 105, L040201 (2022)}}.
\bibitem{Heinosaari} T. Heinosaari and M. Ziman, The Mathematical Language of Quantum Theory: From Uncertainty to Entanglement (Cambridge University Press 2011).
\end{references}
\end{document}